\documentclass[prc,twocolumn,showpacs,superscriptaddress, twoside,showpacs]{revtex4}
\usepackage{graphicx}
\usepackage{dcolumn}
\usepackage{color}
\usepackage{bm}
\usepackage{CJK}
\usepackage{amsmath}
\allowdisplaybreaks[4]

\begin{document}
\begin{CJK*}{GBK}{song}

\title{Tidal wave in $^{102}$Pd:  An extended five-dimensional collective Hamiltonian description}
\author{Y. Y. Wang}
 \affiliation{School of Physics and Nuclear Energy Engineering and International Research Center
 for Nuclei and Particles in the Cosmos, Beihang University, Beijing 100191, China}

\author{Z. Shi}
 \affiliation{School of Physics and Nuclear Energy Engineering and International Research Center
 for Nuclei and Particles in the Cosmos, Beihang University, Beijing 100191, China}

\author{Q. B. Chen }
 \affiliation{State Key Laboratory of Nuclear Physics and Technology, School of Physics, Peking
University, Beijing 100871, China}

\author{S. Q. Zhang }\email{sqzhang@pku.edu.cn}
 \affiliation{State Key Laboratory of Nuclear Physics and Technology, School of Physics, Peking
University, Beijing 100871, China}

\author{C. Y. Song}\email{cysong@buaa.edu.cn}
 \affiliation{School of Physics and Nuclear Energy Engineering and International Research Center
 for Nuclei and Particles in the Cosmos, Beihang University, Beijing 100191, China}

\date{\today}
\begin{abstract}
  The five-dimensional collective Hamiltonian based on the covariant density functional theory
  is applied to investigate the observed tidal wave mode in the yrast band of $^{102}$Pd.
  The energy spectra, the relations between the spin and the rotational frequency, and the
  ratios of $B(E2)/\mathcal{J}(I)$ in the yrast band are well reproduced by introducing the
  empirical $ab$ formula for the moments of inertia. This $ab$ formula is related to the fourth
  order effect of collective momentum in the collective Hamiltonian. It is also shown that the
  shape evolution in the tidal wave is determined microscopically by the competition between the
  rotational kinetic energy and the collective potential in the framework of the collective Hamiltonian.
\end{abstract}
 \maketitle
\date{today}

\section{Introduction}

Over the past two decades, novel nuclear excitation modes, such as the magnetic~\cite{frauendorf1994proceedings,
clark2000shears,frauendorf2001spontaneous,hubel2005magnetic,meng2013progress} and antimagnetic~\cite{frauendorf1995proceedings,
frauendorf2001spontaneous} rotations in near spherical nuclei, and the wobbling~\cite{bohr1975nuclear,odegaard2001evidence}
and chiral~\cite{frauendorf1997tilted,starosta2001chiral,frauendorf2001spontaneous,meng2010open} motions in the triaxially
deformed nuclei, have attracted significant attention and extensively been explored from both the experimental and theoretical
sides. Recently, another novel nuclear excitation mode, named as the ``tidal wave,'' has gradually come into people's vision and
attracts new attention~\cite{pattison2003multiphonon,cullen2005nuclear,frauendorf2011tidal,ayangeakaa2013tidal,frauendorf2015low}.

As illustrated in Ref.~\cite{frauendorf2011tidal}, for a quadrupole vibrating droplet of an ideal liquid, there are a family of
flow patterns with the same energy, differing by their angular momenta. In the family, there are two limits. One limit is the
oscillating motion with no angular momentum carrying and the other is the traveling wave with maximal angular momentum
($2n\hbar$, where $n$ denotes the phonon number) carrying. For a latter case, the surface rotates with a constant angular
velocity as in the case of the rotation of a rigid body. This is the so-called tidal wave mode~\cite{frauendorf2011tidal}.
For a finite nuclear system, this mode is expected to appear in a vibrational or transitional nucleus and corresponds to its
yrast mode.

The concept of the tidal wave is of particular concern since it provides a new mechanism for the generation of the angular momentum.
Compared with a rigid rotor, where the energy and the angular momentum increase with the rotational angular frequency, the energy
and the angular momentum in a tidal wave mode increase with the amplitude of the surface vibration but its frequency remains nearly
constant. Quantally the increase of the amplitude of the tidal wave corresponds to the condensation of the phonons~\cite{frauendorf2011tidal}.
For a quadrupole tidal wave, it is due to the condensation of quadrupole phonons (d-bosons)~\cite{frauendorf2011tidal}, and for an octupole
tidal wave, it is due to octupole phonons (f-bosons)~\cite{frauendorf2013cross}.

Experimentally, the tidal wave was first reported in $^{182}$Os~\cite{pattison2003multiphonon} and then in $^{181,183}$Os~\cite{cullen2005nuclear},
where the spin difference $\Delta I=1$ multiphonon vibration sequences based on high spin intrinsic states were interpreted as tidal waves running over a
triaxial surface. Subsequently, the alternating parity sequences in $^{220}$Th~\cite{reviol2006multiple} and $^{219}$Th~\cite{reviol2009parity}
were regarded as the reflection-asymmetric tidal wave traveling over a spherical core. Very recently, based on the lifetime measurements, the yrast
states of $^{102}$Pd were identified to be the tidal wave running over a quadrupole surface~\cite{ayangeakaa2013tidal}. In this nucleus, the energies
of the yrast band are nearly equidistant up to spin $I=14$, and the extracted experimental reduced transition probabilities $B(E2)$ display a monotonic
increase with spin, which provides the first clear evidence of the seven boson condensation.

Theoretically, the tidal wave mode was first investigated in the framework of the shell correction version of the tilted axis cranking model
(SCTAC)~\cite{frauendorf2011tidal}. Using this model, the observed tidal waves in $^{181,182, 183}$Os~\cite{pattison2003multiphonon,
cullen2005nuclear} and $^{102}$Pd~\cite{ayangeakaa2013tidal} were well reproduced. Later on, a phenomenological phonon model that includes
anharmonic terms was introduced to analyze the rotational and electromagnetic properties of the tidal wave in $^{102}$Pd
\cite{macchiavelli2014tidal}. It is worth noting that the angular momentum in the cranking model is not a good quantum number, and the electromagnetic
transition probabilities could only be treated in a semiclassical manner. Therefore, it is necessary to search for a theoretical model, in which
the angular momentum and the electromagnetic transition probabilities are treated in a quantal manner to investigate the tidal waves.

Since its introduction, the covariant density functional theory (CDFT) has achieved great success in exploring the ground state properties of
both spherical and deformed nuclei over almost the whole nuclide chart~\cite{ring1996relativistic,vretenar2005relativistic,meng2006relativistic,
meng2011covariant,liang2015hidden,meng2015halos} on the basis of relativistic energy density functionals without any additional parameters.
Normally in these studies of ground-states the static nuclear mean field approximation is adopted. To calculate the excitation energies and
electromagnetic transition probabilities of nuclear low-lying spectra, one needs to include the correlations beyond the static mean field through
the restoration of broken symmetries and configuration mixing of symmetry-breaking product states. An effective approach is to construct a
collective Hamiltonian with collective parameters determined from microscopic self-consistent mean-field calculations. Particularly to include
the rotational symmetry restoration and take into account triaxial shape fluctuations, the five-dimensional collective Hamiltonian based on the
covariant density functional theory (5DCH-CDFT) has been developed~\cite{nikvsic2009beyond,nikvsic2011relativistic}. In this model, five quadrupole
dynamical degrees of freedom including the deformation parameters $\beta$ and $\gamma$ and the orientation angles $\Omega=(\phi,\theta,\psi)$
of the nucleus are considered. The broken rotational symmetry in the static mean field is restored by taking all of the possible orientations into
account. Meanwhile, the shape fluctuations around the mean-field minima are allowed by constructing the total Hamiltonian on the quadrupole deformation
space. The 5DCH-CDFT has been extensively applied to describe the nuclear collective properties, such as the phase transitions~\cite{Li2009,li2009microscopic,
li2010microscopic,song2011microscopic,li2013simultaneous}, the shape evolutions~\cite{li2011energy,xiang2012covariant,sun2014spectroscopy,
nikvsic2014microscopic,wang2015covariant} as well as the low-lying spectra along with the isotopic and isotonic chains in different mass regions~\cite{nikvsic2009beyond,li2010relativistic,li2012enhanced,fu2013beyond,mei2012rapid,zshi2015}. For a review, see, e.g., Ref.
\cite{nikvsic2011relativistic}.

Therefore it is interesting to investigate the tidal waves in the vibrational or transitional nuclei by applying the 5DCH-CDFT. On the one hand,
it can answer the question whether the 5DCH could well describe this novel excitation mode or not; on the other hand, it might shed light on the
tidal wave study from the point of view of the collective Hamiltonian. In this paper, taking $^{102}$Pd as an example, the 5DCH-CDFT is applied to
investigate the energy spectra and electromagnetic properties of the tidal wave mode.

The present paper is organized as follows. In Sec.~\ref{sec1}, a brief introduction to the frameworks of the CDFT and the 5DCH is given, together with a comment on the microscopic basis of 5DCH. In Sec.~\ref{sec2}, the potential energy surface of $^{102}$Pd obtained by the CDFT, the yrast energy spectra, and the reduced transition probabilities obtained from the 5DCH in comparison with the experimental data are presented. It is found that
in order to better describe the tidal wave in $^{102}$Pd, it is necessary to introduce a spin dependent moment of inertia into the 5DCH
calculations. By giving the average quadrupole deformation parameters and the probability distributions of collective wave functions, the shape
evolution for the tidal wave in $^{102}$Pd is also analysed. Finally, a summary is given in Sec.~\ref{sec3}.


\section{Theoretical framework}\label{sec1}

In this section, first, a brief introduction to the theoretical framework of the covariant density functional theory is presented. Then the
formalism of the five-dimensional collective Hamiltonian is given together with its derivation from the adiabatic self-consistent collective
coordinate~(ASCC) method. For a simple case with one collective parameter, the derivation is further extended to include the fourth order
of the collective momenta $\bm{p}^4$ in the collective Hamiltonian in the last subsection.

\subsection{Covariant density functional theory}
Detailed formalism of the CDFT could be found in many literatures, such as Refs.~\cite{ring1996relativistic,vretenar2005relativistic,
meng2006relativistic,meng2011covariant,meng2013progress,liang2015hidden,meng2015halos,nikvsic2011relativistic}. The starting point of
the CDFT is a general effective Lagrangian density where the nucleons are coupled with either a meson exchange interaction
\cite{ring1996relativistic,vretenar2005relativistic,meng2006relativistic} or zero-range point-coupling interaction
\cite{nikolaus1992nuclear,Burvenich2002,zhao2010new} as follows:
\begin{align}
  \mathcal{L}&=\mathcal{L}^{\rm free}+\mathcal{L}^{\rm 4f}+\mathcal{L}^{\rm hot}+\mathcal{L}^{\rm der}+\mathcal{L}^{\rm em}\notag\\
             &=\bar{\psi}(i\gamma_{\mu}\partial^{\mu}-m)\psi-\frac{1}{2}\alpha_S(\bar{\psi}\psi)(\bar{\psi}\psi)
              -\frac{1}{2}\alpha_V(\bar{\psi}\gamma_{\mu}\psi)(\bar{\psi}\gamma^{\mu}\psi)\notag\\
             &-\frac{1}{2}\alpha_{TS}(\bar{\psi}\vec{\tau}\psi)(\bar{\psi}\vec{\tau}\psi)
              -\frac{1}{2}\alpha_{TV}(\bar{\psi}\vec{\tau}\gamma_{\mu}\psi)(\bar{\psi}\vec{\tau}\gamma^{\mu}\psi)\notag\\
             &-\frac{1}{3}\beta_{S}(\bar{\psi}\psi)^3-\frac{1}{4}\gamma_{S}(\bar{\psi}\psi)^4
              -\frac{1}{4}\gamma_{V}[(\bar{\psi}\gamma_{\mu}\psi)(\bar{\psi}\gamma^{\mu}\psi)]^2\notag\\
             &-\frac{1}{2}\delta_S\partial_{\nu}(\bar{\psi}\psi)\partial^{\nu}(\bar{\psi}\psi)
              -\frac{1}{2}\delta_V\partial_{\nu}(\bar{\psi}\gamma_{\mu}\psi)\partial^{\nu}(\bar{\psi}\gamma^{\mu}\psi)\notag\\
             &-\frac{1}{2}\delta_{TS}\partial_{\nu}(\bar{\psi}\vec{\tau}\psi)\partial^{\nu}(\bar{\psi}\vec{\tau}\psi)
              -\frac{1}{2}\delta_{TV}\partial_{\nu}(\bar{\psi}\vec{\tau}\gamma_{\mu}\psi)
              \partial^{\nu}(\bar{\psi}\vec{\tau}\gamma_{\mu}\psi)\notag\\
             &-\frac{1}{4}F^{\mu\nu}F_{\mu\nu}-e\frac{1-\tau_3}{2}\bar{\psi}\gamma^{\mu}\psi A_{\mu}.\label{eq06}
\end{align}
In Eq.~(\ref{eq06}), $m$ is the nucleon mass, $e$ the charge unit for protons, and $A_{\mu}$ and $F_{\mu\nu}$, respectively, the four-vector
potential and field strength tensor of the electromagnetic field. For the 11 coupling constants $\alpha_S$, $\alpha_V$, $\alpha_{TS}$,
$\alpha_{TV}$, $\beta_S$, $\gamma_S$, $\gamma_V$, $\delta_S$, $\delta_V$, $\delta_{TS}$, and $\delta_{TV}$, $\alpha$ refers to the
four-fermion terms, $\beta$ and $\gamma$, respectively, the third- and fourth-order terms, and $\delta$ the derivative couplings. The
subscripts $S$, $V$, and $T$ indicate the symmetries of the couplings, i.e., $S$ stands for scalar, $V$ for vector, and $T$
for isovector.

In this work, the relativistic density functional PC-PK1~\cite{zhao2010new} and the density-independent $\delta$-force are, respectively,
adopted in the particle-hole and particle-particle channels, with pairing correlations treated in the Bardeen-Cooper-Schrieffer (BCS)
approximation. The energy density functional for a nuclear system can be self-consistently obtained in terms of local single-nucleon
densities and currents~\cite{zhao2010new}:
\begin{align}
  E_{\rm DF}=&\int d^3r  \mathcal{E}(\bm r)\notag\\
             =&\int d^3r \sum_{k}\nu_k^2\psi_k^\dag(\bm r)(-i\bm{\alpha}\cdot\bm{p}+m)\psi_k(\bm r)\notag\\
              &+\int d^3 r \Bigg(\frac{\alpha_S}{2}\rho_S^2+\frac{\beta_S}{3}\rho_S^3
               +\frac{\gamma_S}{4}\rho_S^4+\frac{\delta_S}{2}\rho_S\Delta\rho_S\notag\\
              &+\frac{\alpha_V}{2}j_{\mu}j^{\mu}+\frac{\gamma_V}{4}(j_{\mu}j^{\mu})^2
               +\frac{\delta_V}{2}j_{\mu}\Delta j^{\mu}\notag\\
              &+\frac{\alpha_{TV}}{2}\vec{j}_{TV}^{\mu}(\vec{j}_{TV})_{\mu}
               +\frac{\delta_{TV}}{2}\vec{j}_{TV}^{\mu}\Delta(\vec{j}_{TV})_{\mu}\notag\\
              &+\frac{1}{4}F_{\mu\nu}F^{\mu\nu}-F^{0\mu}\partial_0A_\mu+eA_\mu j_p^\mu\Bigg), \label{eq07}
\end{align}
in which $\psi$ denotes the Dirac spinor field of a nucleon. The local densities and currents
\begin{align}
  \rho_S(\bm r)&=\sum_k\nu_k^2\bar{\psi}_k(\bm r)\psi_k(\bm r),\\
  j^{\mu}(\bm r)&=\sum_k\nu_k^2\bar{\psi}_k(\bm r)\gamma^{\mu}\psi_k(\bm r),\\
  \vec{j}_{TV}^{\mu}(\bm r)&=\sum_k\nu_k^2\bar{\psi}_k(\bm r)\vec{\tau}\gamma^{\mu}\psi_k(\bm r),\label{eq08}
\end{align}
are calculated in the no-sea approximation, i.e., the summation in Eqs.~(\ref{eq07}) to (\ref{eq08}) only runs over all occupied states with positive
energies, where $\nu_k^2$ represents the occupation factors of single-nucleon states.

By minimizing the energy density functional Eq.~(\ref{eq07}) with respect to $\bar{\psi}_k$, one obtains the Dirac equation for the single nucleon:
\begin{equation}
  [\gamma_{\mu}(i\partial^{\mu}-V^{\mu})-(m+S)]\psi_k=0.\label{eq001}
\end{equation}

To describe nuclei with general quadrupole shapes, the Dirac equation (\ref{eq001}) is solved by expanding the nucleon spinors in the basis of a
three dimensional harmonic oscillator in Cartesian coordinates. The map of the energy surface as a function of the quadrupole deformation is obtained
by imposing constraints on the axial and triaxial quadrupole moments. The method of quadratic constraint uses an unrestricted variation of the function
\begin{align}
  \langle \hat{H}\rangle+\sum_{\mu=0,2}C_{2\mu}(\langle\hat{Q}_{2\mu}\rangle-q_{2\mu})^2,
\end{align}
where $\langle \hat{H}\rangle$ is the total energy, and $\langle\hat{Q}_{2\mu}\rangle$ denote the expectation values of the mass quadrupole operators:
\begin{align}
  \hat{Q}_{20}=2z^2-x^2-y^2~{\rm and}~\hat{Q}_{22}=x^2-y^2.
\end{align}
Here $q_{2\mu}$ are the constrained values of the quadrupole moments, and $C_{2\mu}$ the corresponding stiffness constants~\cite{ring1980nuclear}.

With the single-nucleon wave functions, energies, and occupation factors, generated from constrained self-consistent CDFT solutions, one can
calculate the collective inertia parameters. By using the Inglis-Belyaev (IB) formula~\cite{inglis1956nuclear,beliaev1961concerning},
the moments of inertia~(MOIs) are
\begin{equation}
  \mathcal{I}_k^{\rm IB}=\sum_{ij}\frac{(u_iv_j-v_iu_j)^2}{E_i+E_j}|\langle i|
\hat{J}_k|j\rangle|^2, ~~~~ k=1,2,3, \label{eq05}
\end{equation}
in which $k$ denotes the principal axis of nucleus, and $E_i$ and $E_j$ denote the quasiparticle energy of the quasiparticle states $|i\rangle$
and $|j\rangle$. The summations $i$ and $j$ run over the proton and neutron quasiparticle states. Similar to the MOIs, the mass parameters
associated with the two quadrupole collective coordinates $q_0=\langle\hat{Q}_{20}\rangle$ and $q_2=\langle\hat{Q}_{22}\rangle$ can be also
calculated in the cranking approximation~\cite{girod1979zero}
\begin{equation}
  B_{\mu\nu}(q_0,q_2)=\frac{\hbar^2}{2}[\mathcal{M}_{(1)}^{-1}\mathcal{M}_{(3)}\mathcal{M}_{(1)}^{-1}]_{\mu\nu},
  \label{eq16}
\end{equation}
with
\begin{equation}
  \mathcal{M}_{(n),\mu\nu}(q_0,q_2)=\sum_{ij}
\frac{\langle i|\hat{Q}_{2\mu}|j\rangle\langle j|\hat{Q}_{2\nu}|i\rangle}{(E_i+E_j)^n}(u_iv_j+v_iu_j)^2.
\end{equation}

\subsection{Five-dimensional collective Hamiltonian}

The collective Hamiltonian, in terms of a few numbers of collective coordinates and momenta, is an efficient method for describing various
kinds of large amplitude collective motions, e.g., the shape coexistence~\cite{prochniak2009quadrupole,nikvsic2011relativistic}, the fission
\cite{ring1980nuclear}, and the chiral and the wobbling motions~\cite{chen2013collective,chen2014collective,qbchen2015}. As a phenomenological
model, it can be derived from some microscopic methods~\cite{prochniak2009quadrupole}, e.g., the adiabatic time dependent Hartree-Fock~(ATDHF)
method~\cite{belyaev1965time,baranger1968nuclear,baranger1978adiabatic,ring1980nuclear}, the generate coordinate method~(GCM)
\cite{hill1953nuclear,une1976collective,ring1980nuclear,gozdz1985mass}, and the ASCC method~\cite{matsuo2000adiabatic,hinohara2008microscopic,
hinohara2009microscopic,hinohara2010microscopic,ascc-matsuyanagi2010open}.

In the following, the procedure for the derivation of the collective Hamiltonian with the ASCC method is briefly presented. The main concept of the
ASCC method is to solve the equations of the self-consistent collective coordinate (SCC) method~\cite{SCC-Marumori1977,marumori1980self} using an
expansion with respect to the collective momentum. The starting point of the SCC method is the time dependent Hartree-Fock (TDHF) equation.
By assuming that the TDHF determinantal states could be represented in multi-dimensional classical phase space characterized by a set of
collective coordinates $\bm{p}=\{p_1,p_2,\ldots\}$ and collective momenta $\bm{q}=\{q_1, q_2, \ldots\}$, the collective Hamiltonian is defined
as the expectation value of the nuclear many-body Hamiltonian $\hat{H}$,
\begin{align}
 H_{\rm coll}=\langle\phi(\bm{q},\bm{p})|\hat{H}|\phi(\bm{q},\bm{p})\rangle.
\end{align}
With the adiabatic approximation which assumes the collective motion to be slow compared to single-particle motion in the nucleus, the collective
Hamiltonian is expanded in powers of the collective momentum $\bm{p}$, stopping at the second order,
\begin{equation}\label{eq14}
  H_{\rm coll}=\frac{1}{2}\sum_{ij}B_{ij}(\bm{q})p_ip_j+V(\bm{q}),
\end{equation}
where $B_{ij}(\bm{q})$ and $V(\bm{q})$ are the so-called mass parameter and collective potential, respectively. In the framework of the ASCC method, these two
quantities are self-consistently obtained by solving the ASCC equations including the moving-frame Hartree-Fock-Bogoliubov (HFB) equation and
moving-frame Random Phase Approximation (RPA) equations~\cite{matsuo2000adiabatic,hinohara2008microscopic,hinohara2009microscopic,
hinohara2010microscopic,ascc-matsuyanagi2010open}. Here, we would not repeat the corresponding formulas again.

When the ASCC method is applied to describe the collective rotation, vibration, and the couplings between them, the well-known Bohr Hamiltonian
\cite{bohr1975nuclear}, or referred to as the 5DCH, can be constructed in terms of the five collective intrinsic variables $\beta$, $\gamma$, and Euler
angles $\Omega=\{\phi,\theta,\psi\}$. The 5DCH is written as
 \begin{equation}
  H_{\rm coll}=T_{\rm vib}(\beta, \gamma)
   +T_{\rm rot}(\beta, \gamma, \Omega)+V(\beta, \gamma),\label{eq01}
 \end{equation}
with the vibrational kinetic energy
 \begin{equation}
  T_{\rm vib}=\frac{1}{2}B_{\beta\beta}\dot{\beta}^2+\beta B_{\beta\gamma}\dot{\beta}\dot{\gamma}
  +\frac{1}{2}\beta^2B_{\gamma\gamma}\dot{\gamma}^2,
 \end{equation}
the rotational kinetic energy
\begin{equation}
 T_{\rm rot}=\frac{1}{2}\sum_{k=1}^{3}\mathcal{I}_k\omega_k^2,
\end{equation}
and the collective potential energy $V$. The mass parameters $B_{\beta\beta}$, $B_{\beta\gamma}$, $B_{\gamma\gamma}$, and the
MOIs $\mathcal{I}_k$ depend on the collective variables $\beta$ and $\gamma$.

The Hamiltonian (\ref{eq01}) is quantized according to the Pauli prescription~\cite{pauli1933handbuch} as
 \begin{equation}
  \hat{H}_{\rm coll}=\hat{T}_{\rm vib}+\hat{T}_{\rm rot}+V,\label{eq10}
 \end{equation}
with
 \begin{align}
 \hat{T}_{\rm vib}=&-\frac{\hbar^2}{2\sqrt{wr}}\Bigg\{\frac{1}{\beta^4}\Bigg[\frac{\partial}{\partial\beta}
                     \sqrt{\frac{r}{w}}\beta^4B_{\gamma\gamma}\frac{\partial}{\partial\beta}\notag\\
                   &-\frac{\partial}{\partial\beta}\sqrt{\frac{r}{w}}\beta^3B_{\beta\gamma}\frac{\partial}
                       {\partial\gamma}\Bigg]+\frac{1}{\beta\sin{3\gamma}}\Bigg[-\frac{\partial}{\partial\gamma}\notag\\
                   &\times\sqrt{\frac{r}{w}}\sin{3\gamma}B_{\beta\gamma}
                     \frac{\partial}{\partial\beta}+\frac{1}{\beta}\frac{\partial}
                     {\partial\gamma}\sqrt{\frac{r}{w}}\sin{3\gamma}B_{\beta\beta}
                     \frac{\partial}{\partial\gamma}\Bigg]\Bigg\},\label{eqTvib}
 \end{align}
and
\begin{align}
  \hat{T}_{\rm rot}=\frac{1}{2}\sum_{k=1}^{3}\frac{\hat{J}_k^2}{\mathcal{I}_k},\label{eqTrot}
\end{align}
where $\hat{J}_k$ denotes the components of the angular momentum in the body-fixed frame of a nucleus. The two quantities that appear in the
expression~(\ref{eqTvib}) for the vibrational energy are $r=D_1D_2D_3$ and $w=B_{\beta\beta}B_{\gamma\gamma}-B^2_{\beta\gamma}$, where the
inertial parameters $D_k$ are related to the MOIs $\mathcal{I}_k$ as,
\begin{equation}
  \mathcal{I}_k=4D_k\beta^2\sin^2\left(\gamma-\frac{2k\pi}{3}\right),~k=1,2,3.
\end{equation}

The corresponding eigenvalue problem is solved using an expansion of eigenfunctions in terms of a complete set of basis functions that depend
on the deformation variables $\beta$ and $\gamma$ and the Euler angles $\phi$, $\theta$, and $\psi$~\cite{Prochniak1999}. The diagonalization
of the Hamiltonian yields the excitation energies and collective wave functions:
\begin{equation}
  \Psi_{\alpha}^{IM}(\beta, \gamma, \bm{\Omega})=\sum_{K\in \Delta I}\psi_{\alpha K}^I(\beta,
  \gamma)\Phi_{MK}^I(\bm{\Omega}).\label{eq09}
\end{equation}
where the summation is over the allowed set of $K$ values:
\begin{align}
 \Delta I=
 \begin{cases}
  0, 2, \dots, I    & {\rm for}  ~I~ {\rm mod}~ 2 = 0\\
  2, 4, \dots, I-1  & {\rm for}  ~I~ {\rm mod}~ 2 = 1.
 \end{cases}
\end{align}

Using the collective wave functions ({\ref{eq09}}), various observables can be calculated and compared with experimental data. For the quadrupole
$E2$ reduced transition probability, it is calculated as
\begin{equation}\label{eq13}
 B(E2: \alpha I \rightarrow \alpha' I')=\frac{1}{2I+1}|\langle\alpha' I'||\hat{\mathcal{M}}(E2)||\alpha I\rangle|^2,
\end{equation}
where $\hat{\mathcal{M}}(E2)$ denotes the electric quadrupole operator; see Ref.~\cite{kumar1967complete} for details.

In the framework of 5DCH-CDFT, all the collective parameters are determined by the CDFT calculations. That is, the collective potential energy
$V$ is given by the constrained CDFT calculations in the $\beta$-$\gamma$ plane; meanwhile the moments of inertia $\mathcal{I}_k$
and the mass parameters $B_{\beta\beta}$, $B_{\beta\gamma}$, $B_{\gamma\gamma}$ may be approximately given by the cranking formulas~(\ref{eq05})
and (\ref{eq16}), respectively.

\subsection{A comment on expansions in the collective momentum}\label{sec4}
As stated in the last subsection, the collective Hamiltonian can be derived from the ASCC method by expanding the collective momenta up to the second order.
It should be noticed that, as commented on in Ref.~\cite{baranger1978adiabatic}, the adiabatic assumption is actually vital to the phenomenological
forms of the collective model, since the kinetic energy would not be quadratic otherwise, except in the trivial case of translations. However,
there is no reason why the energy in the collective model should turn out to be a quadratic function of the collective momenta or the velocities in general.
The only way to get it to be quadratic is by assuming that the velocities are small and by expanding the energy in powers of them, stopping at the second order~\cite{baranger1978adiabatic}.

Although the collective Hamiltonian with this assumption has achieved lots of successes, it was found that the collective Hamiltonian could not well
describe the energy spectra for the weakly deformed transitional nuclei when the spin is large~\cite{nikvsic2009beyond}. This fact might suggest that
the expansion of the collective Hamiltonian in collective momentum up to the second order is not necessarily enough. It would be therefore interesting to
study the contributions of the higher-order terms of the collective momenta.

As an initial try, we take the fourth order of the collective momentum $\bm{p}^4$ into account in the collective Hamiltonian. For simplicity, we
consider the one dimensional case; then the Eq.~(\ref{eq14}) becomes
\begin{align}\label{eq15}
 H_{\rm coll}=\frac{1}{2!}B_{2}(q)p^2+\frac{1}{4!}B_{4}(q)p^4+V(q).
\end{align}
Following the procedure of deriving the ASCC equations in Ref.~\cite{matsuo2000adiabatic}, the mass parameters $B_2(q)$, $B_4(q)$, and the collective
potential $V(q)$ in Eq.~(\ref{eq15}) are now calculated by
\begin{align}
   V(q)&=\langle\phi(q)|\hat{H}|\phi(q)\rangle,\\
   B_{2}(q)&=\frac{\partial^2H_{\rm coll}}{\partial q^2}\Bigg|_{p=0}
                       =-\langle\phi(q)|\hat{Q},[\hat{Q},\hat{H}]]|\phi(q)\rangle,\\
   B_{4}(q)&=\frac{\partial^4H_{\rm coll}}{\partial q^4}\Bigg|_{p=0}
                       =\langle\phi(q)|[\hat{Q},[\hat{Q},[\hat{Q},
                       [\hat{Q},\hat{H}]]]]|\phi(q)\rangle,
\end{align}
where $\hat{Q}$ is the infinitesimal generator defined at the collective state $|\phi(q)\rangle$
\begin{align}
  \hat{Q}|\phi(q)\rangle&=\frac{1}{i}\frac{\partial|\phi(q,p)\rangle}{\partial p}\Bigg|_{p=0}.
\end{align}
If we neglect the residual two-body interaction in the Hamiltonian $\hat{H}$, the mass parameters are reduced to the cranking formulas,
\begin{align}
      B_{2}^{-1}(q)&=2\sum_{mi}\frac{|\langle m|\frac{\partial \hat{H}_M}{\partial q}|i\rangle|^2}
      {(\varepsilon_{m}-\varepsilon_{i})^3},\\
      B_{4}^{-1}(q)&=8\sum_{mi}\frac{|\langle m|\frac{\partial\hat{H}_M}{\partial q}|i\rangle|^4}
      {(\varepsilon_{i}-\varepsilon_{m})^5},
\end{align}
where $\hat{H}_M$ is the so-called moving frame Hamiltonian~\cite{matsuo2000adiabatic}
\begin{align}
 \hat{H}_M(q)=\hat{H}-\frac{\partial V(q)}{\partial q}\hat{Q},
\end{align}
and the $|m\rangle$, $|i\rangle$ and $\varepsilon_{m}$, $\varepsilon_{i}$ are the eigenstates and the corresponding eigenvalues of $\hat{H}_M$.
The indices $m$ and $i$ denote the particle and hole states, respectively. Noting that $\varepsilon_{m}>\varepsilon_{i}$, thus the $B_{2}^{-1}(q)$
is positive, while $B_{4}^{-1}(q)$ is negative.

If one rewrites the collective Hamiltonian (\ref{eq15}) as
\begin{align}
  H_{\rm coll}&=V(q)+\frac{1}{2!}B_2(q)p^2+\frac{1}{4!}B_4(q)p^4\notag\\
             &=V(q)+\frac{1}{2}B_2(q)\left[1+\frac{1}{12}\frac{B_4(q)}{B_2(q)}p^2\right]p^2\notag\\
             &=V(q)+\frac{1}{2}B(q,~p)p^2,
\end{align}
it can be seen that the contributions of the fourth order of $p$ could be absorbed into a momentum dependent effective parameter
\begin{align}
 B(q,~p)=B_2(q)\left[1+\frac{1}{12}\frac{B_4(q)}{B_2(q)}p^2\right].
\end{align}
According to the above analysis,  $B_{2}(q)$ is positive and $B_{4}(q)$ is negative; therefore $B(q,~p)$ should decrease with respect to the increase of $p$.

For considering the fourth order contribution into a rotational kinetic energy with the expression,
\begin{align}
 \hat{T}_{\rm rot}=\frac{\hat{J}^2}{2\mathcal{I}},
\end{align}
one just needs to replace the $p$ with $\hat{J}$, and $1/\mathcal{I}$ with
\begin{align}
 \frac{1}{\mathcal{I}}=B(I)=B_2\left[1+\frac{1}{12}\frac{B_4}{B_2}I(I+1)\right].
\end{align}
In the above expression, the $\hat{J}^2$ has been replaced by $I(I+1)$ in the mean-field approximation. Considering that $B_4$ should be generally far
smaller than $B_2$, $B_4/B_2$ is a small quantity, so that it has
\begin{align}
  \frac{1}{\mathcal{I}}=B_2\Big[1+\frac{1}{12}\frac{B_4}{B_2}I(I+1)\Big]\approx
  \frac{1}{a\sqrt{1+bI(I+1)}},
\end{align}
with
\begin{align}
 a=\frac{1}{B_2}, \quad b=-\frac{B_4}{6B_2}.
\end{align}
This immediately shows that the moment of inertia has the form of
\begin{align}
 \mathcal{I}=a\sqrt{1+bI(I+1)}.
\end{align}
It is just the well-known $ab$ formula proposed by Wu and Zeng~\cite{Wu1985,chong1987new}. From this point of view, the $ab$ formula of MOI can be
microscopically accessed by considering the fourth order term of the collective momentum.
\section{Results and discussion}\label{sec2}

In the CDFT calculations, the point-coupling energy density functional PC-PK1 in the particle-hole channel and the density-independent $\delta$ force
in the particle-particle channel are adopted~\cite{zhao2010new}. The solution of the equation of motion for the nucleons is accomplished by an expansion
of the Dirac spinors in a set of three-dimensional harmonic oscillator basis functions in Cartesian coordinates with 12 major shells. To provide the
collective parameters on the $(\beta,\gamma)$ plane for the 5DCH, a constrained triaxial CDFT calculation is carried out in the region $\beta\in[0.0, 0.8]$
and $\gamma \in[0^{\circ},60^{\circ}]$ with step sizes $\Delta \beta=0.05$ and $\Delta\gamma=6^{\circ}$.

\begin{figure}[htbp]
\centerline{
\includegraphics[scale=0.45,angle=0]{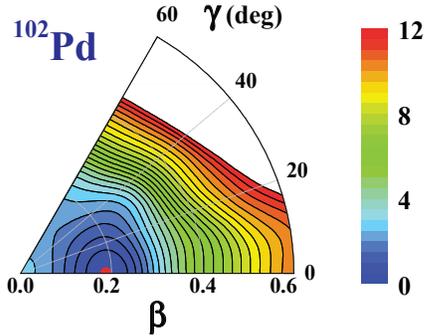}
} \caption{(Color online) The potential energy surface in the $\beta$-$\gamma$ plane
  ($0\leq\gamma\leq60^{\circ}$) for $^{102}$Pd calculated by the constrained triaxial CDFT with PC-PK1~\cite{zhao2010new}. All energies are normalized with
  respect to the binding energy of the absolute minimum (in MeV). The energy between
  each contour line is 0.5 MeV.}
\label{fig1}
\end{figure}

In Fig.~\ref{fig1}, the potential energy surface in the $\beta$-$\gamma$ plane for $^{102}$Pd calculated by the constrained triaxial CDFT is shown. It
shows that the minimum of the potential energy surface (PES) (labeled as a red dot) locates at $(\beta=0.19, \gamma=0^{\circ})$, which corresponds to a moderate prolate shape. Around
the minimum, the PES exhibits a relatively soft character. The energy difference between the ground state $(\beta=0.19,\gamma=0^\circ)$ and the lowest
oblate energy  position $(\beta=0.15,\gamma=60^\circ)$ is less than 2.2 MeV.

With the collective potential, moments of inertia, and mass parameters determined from the CDFT, the collective energies and collective wave functions can be
obtained by diagonalizing the 5DCH at each given spin. It is noted that the  MOIs adopted here are obtained by the IB formula (\ref{eq05}).
It usually underestimates the experimental MOIs due to the absence of the contributions of time-odd fields (the so-called Thouless-Valatin (TV) dynamical
rearrangement contributions)~\cite{nikvsic2009beyond}, the inclusion of which requires much demanding computations. To consider the effect of TV terms,
one may multiply a factor $1+\alpha$ to the IB MOIs, i.e., the input MOIs for the 5DCH are $\mathcal{I}_k=\mathcal{I}_k^{\rm IB}(1+\alpha), k=1,2,3$
\cite{libert1999microscopic,nikvsic2009beyond}. The value of $\alpha$ is obtained so that the energy of the $2_1^+$ state from the 5DCH coincides
with the experimental data. The justification of this treatment has been demonstrated for some nuclei in the $A\sim190$ mass region~\cite{libert1999microscopic},
while the situation may become considerably complicated especially for nuclei with soft potential energy surfaces~\cite{prochniak2004self,prochniak2009quadrupole,
hinohara2010microscopic,hinohara2012effect}.

\begin{figure}
\centering
\includegraphics[scale=0.32,angle=0]{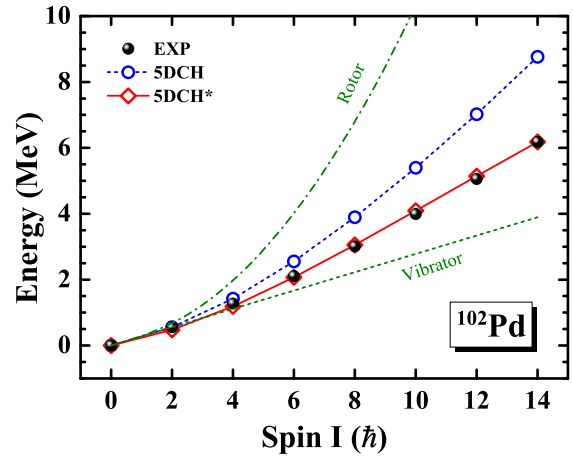}
\caption{(Color online) The comparison of the yrast band energy between the experimental data
and 5DCH calculations for $^{102}$Pd. The blue circle with the dashed line (5DCH) and the red lozenge with the solid line (5DCH*) represent the calculations with the MOI corrected
by renormalizing the IB effective MOI to the empirical values with a common factor and
by the empirical $ab$ formula, respectively. The harmonic vibrational and rotational limits are shown in dashed and dashed-dotted lines.} \label{fig2}
\end{figure}

The obtained energy spectra of the yrast band of $^{102}$Pd from the 5DCH (labeled as 5DCH) in comparison with the data~\cite{ayangeakaa2013tidal} are
illustrated in Fig.~\ref{fig2}.
For reference, the harmonic vibrational and rotational limits are also shown in the figure. It is shown that the experimental data are closer to the harmonic than the rotational limit, and the deviation from the harmonic limit indicates the
anharmonicity as discussed in Refs.~\cite{ayangeakaa2013tidal,macchiavelli2014tidal}.
In the 5DCH calculation, the value of $\alpha$ is taken as 0.40. As shown in Fig.~\ref{fig2}, the 5DCH calculation
can reproduce the data for the low spin region ($I\leq 6\hbar$). However, with the increase of spin, it gradually deviates from the data in value and
also from the linear increase of the data. The deviation reaches to 2.5 MeV at $I=14\hbar$, nearly 40$\%$ larger than the data. This indicates that the
spin-independent MOIs could not describe the energy spectra well for the whole spin region in $^{102}$Pd, whose PES shows a relatively soft character as seen
in Fig.~\ref{fig1}.

As discussed in Sec.~\ref{sec4}, the 5DCH is expanded up to the second order with respect to the collective momentum by using an adiabatic
approximation. It is not necessarily enough for the study of the transitional nuclei. It is therefore interesting to investigate the high order effects
of collective momenta. As shown in Sec.~\ref{sec4}, effectively one can take the fourth order of the collective momentum $p^4$ into account through a $p$
dependent inertial parameter ($\sim p^2$). Correspondingly in the present investigation, we introduce a spin dependent MOIs,
$\mathcal{I}_k=a\sqrt{1+bI(I+1)}=a_0\mathcal{I}_k^{\rm IB}\sqrt{1+bI(I+1)}$, into 5DCH calculations. The parameters $a_0$ and $b$ are obtained by fitting the available
experimental energy spectra using the least squares method (LSM) with $a_0=1.713$ and $b=0.003$. We will denote this calculation as the 5DCH$^*$.

The energy spectra of the 5DCH$^*$ are also included in Fig.~\ref{fig2} in comparison with the 5DCH results and the data. It is seen that 5DCH$^*$ results
well reproduce the data and the maximal difference is only about 0.1 MeV. In addition, it is demonstrated that the energy spectra show a nearly linear
increase with respect to spin. This corresponds to the characteristic energy spectra of a tidal wave~\cite{frauendorf2011tidal}.

\begin{figure}[h!]
\centerline{
\includegraphics[scale=0.32,angle=0]{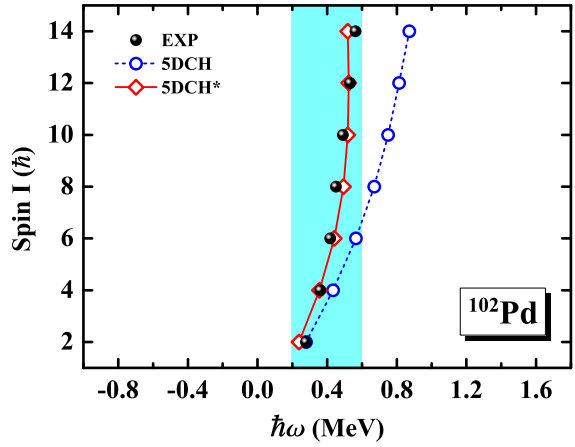}
  } \caption{(Color online) The experimental and theoretical $I$-$\hbar\omega$ relation of the yrast
  band for $^{102}$Pd. The light cyan band represents a region of $\hbar\omega=0.2$ to $0.6$ MeV.} \label{fig3}
\end{figure}

From the energy spectra, the rotational frequency can be extracted according to the classical relations $\omega=dE/dI$ as
\cite{frauendorf1996interpretation}
\begin{align}
  \omega(I)=\frac{\Delta E}{\Delta I}=\frac{1}{2}[E(I)-E(I-2)].
\end{align}
The obtained experimental and theoretical $I$-$\hbar\omega$ relations of the yrast band for $^{102}$Pd are shown in Fig.~\ref{fig3}. It is
seen that the spin increases rapidly with respect to the rotational frequency experimentally. In a very small rotational frequency interval
($\sim 0.4~\rm{MeV}$), the spin increases $12\hbar$. This implies that the generation of spin is mainly caused by the increase of MOIs rather
than the increase of rotational frequency, a significant feature of a tidal wave. The 5DCH$^*$ calculation describes this feature very well.
The 5DCH results however show a slow $I$-$\hbar\omega$ increase, where $\hbar\omega$ increases from $\sim 0.25$ to $0.85~\rm{MeV}$
with spin from $2\hbar$ to $14\hbar$.

\begin{figure}
\centerline{
\includegraphics[scale=0.32,angle=0]{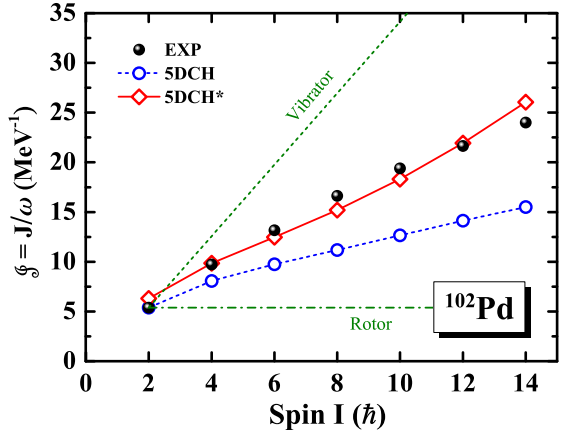}
  } \caption{(Color online) The experimental and theoretical moments of inertia $\mathcal{J}$
  as functions of the spin $I$ for $^{102}$Pd. The harmonic vibrational and rotational limits are shown in dashed and dashed-dotted lines.
  } \label{fig4}
\end{figure}
Figure~\ref{fig4} displays the comparison between the experimental and theoretical moments of inertia $\mathcal{J}$, obtained by $\mathcal{J}=J/\omega$ with
\begin{align}
  \mathcal{J}(I)=\frac{J(I)}{\omega(I)}=\frac{2J(I)}{E(I)-E(I-2)},~~J(I)=I-\frac{1}{2},
\end{align}
in which the classical angular momentum $J$ is associated with the quantal value $(I-1)+1/2=I-1/2$~\cite{frauendorf1996interpretation}. As seen in Fig.~\ref{fig4},
the experimental MOI increases nearly linearly with respect to spin. The 5DCH results give an increasing trend of the MOI but underestimate the increase of
experimental values. With the spin dependent MOIs, the 5DCH$^*$ reproduces the data rather well.
\begin{figure}
\centerline{
\includegraphics[scale=0.55,angle=0]{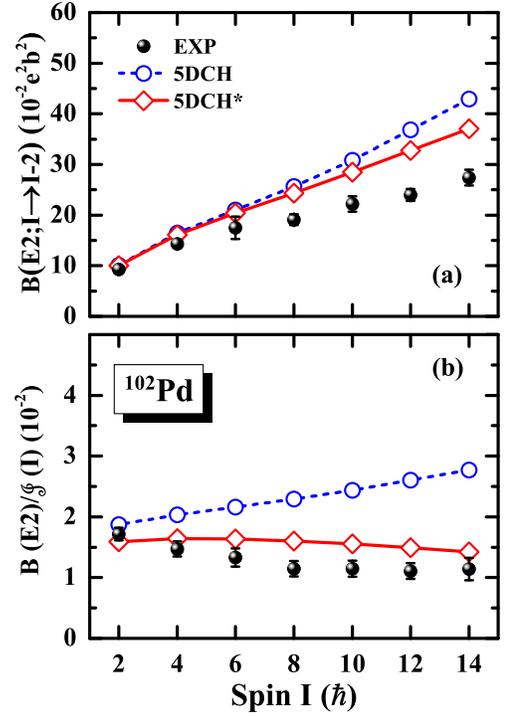}
} \caption{(Color online) (a): The experimental and theoretical reduced electromagnetic transition
  probabilities $B(E2)$ of the yrast band as a function of the spin $I$ for $^{102}$Pd.
  (b): The experimental
  and calculated ratios $B(E2)/\mathcal{J}$  as a function of the spin $I$.
} \label{fig5}
\end{figure}
With the collective wave functions obtained by diagonalizing the 5DCH, the reduced transition
probabilities $B(E2)$ can be calculated according to Eq.~(\ref{eq13}). The corresponding $B(E2)$ values
from the 5DCH and 5DCH$^*$ are drawn in Fig.~\ref{fig5}(a) in comparison with the available data~\cite{ayangeakaa2013tidal}.
It is shown that both the 5DCH and 5DCH$^*$ results correctly give the increase trend of the experimental data,
but to some extent overestimate the experimental values. Again the 5DCH$^*$ results are closer to the data than
those from the 5DCH. It is noted that the monotonic increase of $B(E2)$ values with spin is another characteristic
feature of a tidal wave as demonstrated in Ref.~\cite{frauendorf2011tidal}, which mainly comes from the increase
of the nuclear quadrupole deformation. Considering that in the 5DCH and 5DCH$^*$ calculations the physical observables,
such as transition probabilities and spectroscopic quadrupole moments, are calculated in the full configuration space
and there are no effective charges, such predictions are appreciable. In Fig.~\ref{fig5}(b), the experimental and calculated
ratios $B(E2)/\mathcal{J}$ as functions of spin $I$ are shown, which are expected to be nearly constant for a tidal wave.
This behavior is well shown by the experimental data and also the 5DCH$^*$ results. The 5DCH results, however, give an increase
trend of $B(E2)/\mathcal{J}$ due to the faster increase of the $B(E2)$ with spin than that of MOI. In fact, the behaviors
of $B(E2)$ and $B(E2)/\mathcal{J}$ given by the 5DCH and 5DCH$^*$ can be understood by the evolution of obtained deformation
parameters with respect to spin, which will be discussed below.

\begin{figure}[t!]
\includegraphics[scale=0.52,angle=0]{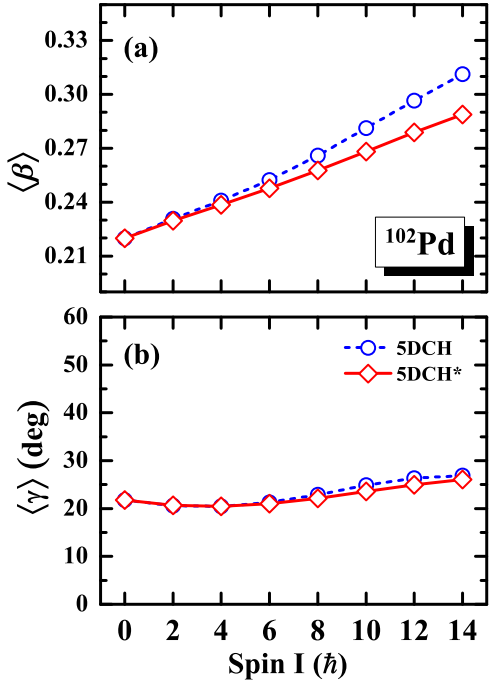}
 \caption{(Color online): The average values of the quadrupole deformation parameters (a) $\beta$ and
  (b) $\gamma$  calculated by the 5DCH as functions of spin $I$ for $^{102}$Pd.
}\label{fig6}
\end{figure}
\begin{figure*}[ht!]
\centering
\includegraphics[scale=0.5, angle=0]{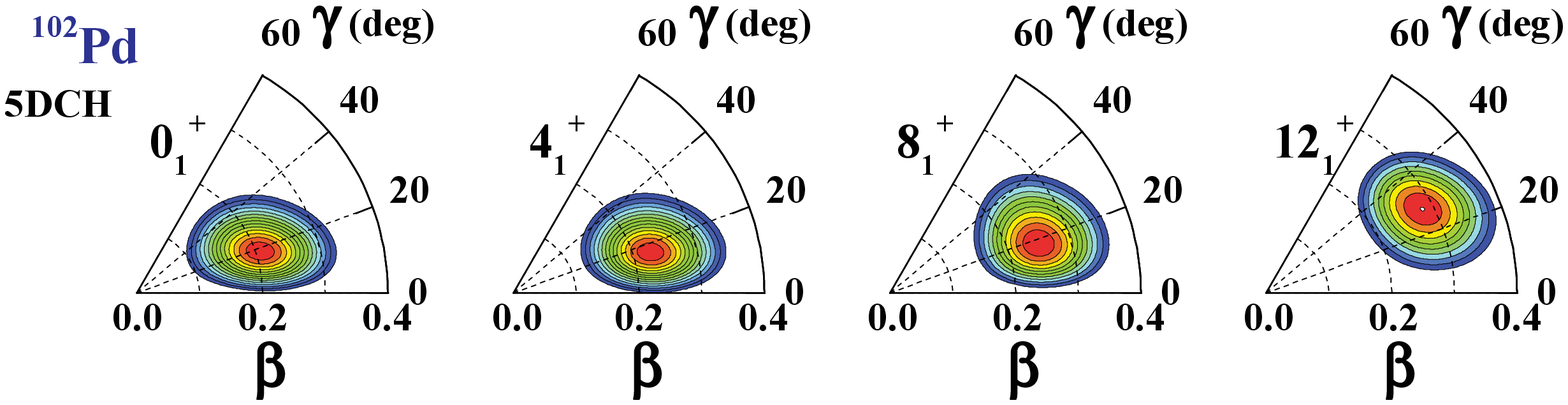} \\
\includegraphics[scale=0.5, angle=0]{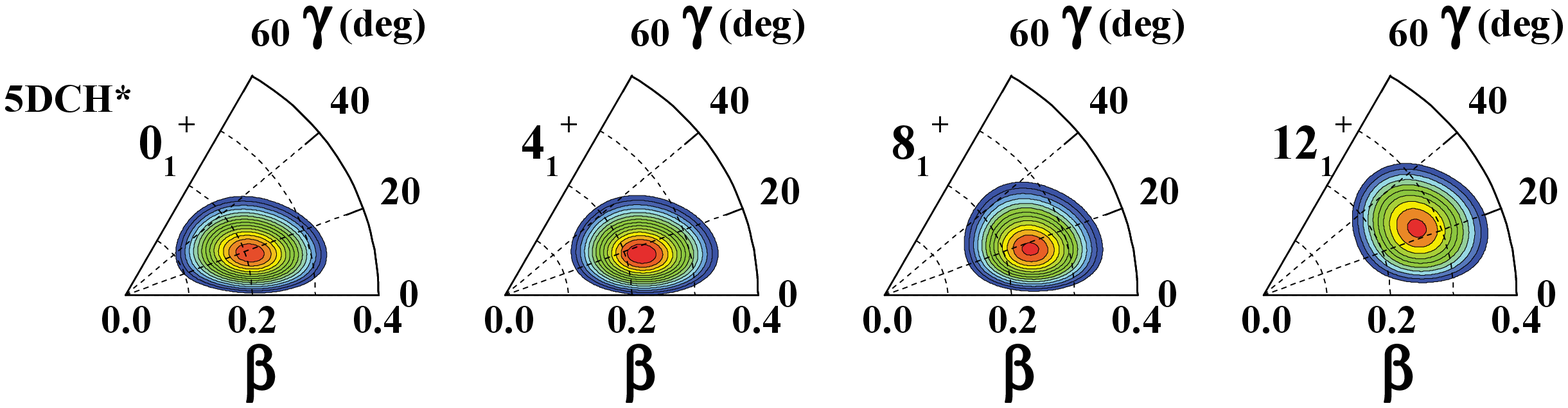}
 \caption{(Color online) The probability distributions in the $\beta$-$\gamma$ plane
 for the $0_1^+$, $4_1^+$, $8_1^+$ and $12_1^+$ states of $^{102}$Pd calculated by the 5DCH
 and 5DCH$^*$.}\label{fig7}
\end{figure*}

In the 5DCH, the expectation values of the quadrupole deformation $\langle\beta\rangle$ and $\langle\gamma\rangle$ for each given state
$|\Psi^I_\alpha\rangle$ are calculated by~\cite{nikvsic2009beyond}
\begin{align}
&\langle\beta\rangle_{I\alpha}=\sqrt{\langle\beta^2\rangle_{I\alpha}},\label{eq11}\\
&\langle\gamma\rangle_{I\alpha}=\frac{1}{3}\arccos\frac{\langle\beta^3\cos3\gamma
\rangle_{I\alpha}}{\sqrt{\langle\beta^2\rangle_{I\alpha}\langle\beta^4\rangle_{I\alpha}}},\label{eq12}
\end{align}
where
\begin{align}
    \langle\beta^2\rangle_{I\alpha}&=\langle\Psi^I_\alpha|\beta^2|\Psi^I_\alpha\rangle\notag\\
    &=\sum_{K\in\Delta I}\int\beta^2|\psi^I_{\alpha,K}(\beta,\gamma)|^2d\tau_0,\\
    \langle\beta^4\rangle_{I\alpha}&=\langle\Psi^I_\alpha|\beta^4|\Psi^I_\alpha\rangle\notag\\
    &=\sum_{K\in\Delta I}\int\beta^4|\psi^I_{\alpha,K}(\beta,\gamma)|^2d\tau_0,\\
    \langle\beta^3\cos3\gamma\rangle_{I\alpha}&=\langle\Psi^I_\alpha|\beta^3\cos3\gamma|\Psi^I_\alpha\rangle\notag\\
    &=\sum_{K\in\Delta I}\int\beta^3\cos3\gamma|\psi^I_{\alpha,K}(\beta,\gamma)|^2d\tau_0.
\end{align}
The calculated $\langle\beta\rangle$ and $\langle\gamma\rangle$ as functions of spin $I$ by the 
5DCH and 5DCH$^*$ are shown in Figs.~\ref{fig6}(a) and \ref{fig6}(b), respectively. For the $\beta$
degree of freedom, the average values $\langle\beta\rangle$ calculated by the 5DCH and 5DCH$^*$
both increase linearly with spin, i.e., from 0.22 at $I=0\hbar$ to 0.31 and 0.29 at $I=14\hbar$, 
respectively, consistent with the increase behavior of $B(E2)$ shown in Fig.~\ref{fig5}(a).
For the $\gamma$ degree of freedom, the average values $\langle\gamma\rangle$ exhibit slight
increases for both the 5DCH and 5DCH$^*$ calculations, i.e., from $\sim22^\circ$ at $I=0\hbar$ to
$\sim27^\circ$ and $\sim26^\circ$ at $I=14\hbar$, respectively.

It is interesting to note that the shape evolution obtained by the 5DCH is more obvious than that by
the 5DCH$^*$. This can be understood by the formalism of the quantized 5DCH~(\ref{eq10}),
the diagonalization of which naturally leads to the collective states at a given angular momentum
and the corresponding distributions in the $\beta$-$\gamma$ plane. Since the vibration term
$\hat{T}_{\rm vib}$ is mainly responsible for the $\beta$ and $\gamma$ vibrational excitations,
for the yrast band focused here, it can be neglected compared with the other two terms $\hat{T}_{\rm rot}$
and $V$. Therefore, the shape evolution of the yrast band is mainly determined by the
competition between the rotational kinetic energy $\hat{T}_{\rm rot}$ and the collective potential
$V$.
In the present 5DCH and 5DCH$^*$ calculations, the collective potential $V$ does not change
with respect to spin, while the rotational kinetic energy $\hat{T}_{\rm rot}$ depends on the quadratic
of spin $I(I+1)$ and also the variation of MOIs.

At the beginning of the yrast band, since the angular momentum is small, its equilibrium  deformation
is mainly determined by the minimum of the collective potential. This explains that the deformation
parameters at the region of $I\leq 4\hbar$ ($\beta\sim 0.22$) are rather close to the deformation
parameter of the minimum of the potential energy surface in Fig.~\ref{fig1} ($\beta\sim 0.20$).
As the spin increases, the contribution of the kinetic energy becomes gradually important. As it is
known, the MOIs increase with respect to $\beta^2$~\cite{ring1980nuclear}, i.e., a larger deformation
$\beta$ would make the kinetic energy more energetic favorable. As a result, the increase of spin
induces the deformation of the yrast state to be larger. This explains the increase trend of the
deformation in the 5DCH as shown in Fig.~\ref{fig6}. After introducing the $ab$ formula in the 5DCH$^*$,
the input MOIs are enlarged. Accordingly, the driven effect by the kinetic energy to the deformation is
reduced. Thus the deformation parameters obtained by 5DCH$^*$ is smaller than the 5DCH ones.

Based on the above analysis, the description of the 5DCH and 5DCH$^*$ for the yrast band of $^{102}$Pd
raises a schematic picture of a tidal wave. Here, the performance of the tidal wave mainly depends on the
competition between the rotational kinetic energy $\hat{T}_{\rm rot}$ and the collective potential $V$.
For a nucleus with a stiff potential energy surface, its deformation is not easy to change and the MOIs are
nearly spin-independent. Thus the increase of the angular frequency will take charge of the generation
of nuclear angular momentum. This corresponds to the rotation of a rigid rotor. For a nucleus with a soft
potential energy surface, its deformation is easy to change and the competition between
$\hat{T}_{\rm rot}$ and $V$ prefers a larger deformation. Thus the increase of the MOIs contributes to
the generation of nuclear angular momentum while the angular frequency
may not have an obvious increase. This corresponds to the mode of a tidal wave.

Finally, in Fig.~\ref{fig7}, the probability distributions~\cite{li2010microscopic} in the $\beta$-$\gamma$ plane
calculated by the 5DCH and 5DCH* are displayed for the yrast states $0_1^+$, $4_1^+$, $8_1^+$, and $12_1^+$ of $^{102}$Pd.
It can be clearly seen that the peaks of probability distributions locate around the deformed points which are consistent
with the values of the average deformation parameters shown in Fig.~\ref{fig6}. To further compare the difference between
the probability distributions from the 5DCH and 5DCH$^*$ calculations is interesting. For low spin states such as $0^+_1$
and $4^+_1$, the 5DCH and 5DCH$^*$ calculations present a similar probability distributions. For higher states such as $8^+_1$
and $12^+_1$, due to the reduction of the kinetic energy driven effect mentioned above, the 5DCH$^*$ calculations present
smaller average deformation $\langle\beta\rangle$ and more concentrated probability distributions than those of 5DCH.

\section{Summary}\label{sec3}

In summary, the five-dimensional collective Hamiltonian based on the covariant density functional theory has been first
applied to investigate the phenomenon of the tidal wave in the yrast band of $^{102}$Pd. Although the 5DCH calculation
with spin-independent MOIs could qualitatively describe the tendency of the yrast band, it overestimates the increase of excitation
energies (in turn underestimates the increase of MOIs) and the ratios of $B(E2)/\mathcal{J}(I)$. Considering that the
adiabatic approximation adopted for the collective momenta up to the second order in the 5DCH may be not enough for the soft
nucleus when the spin is large, the fourth order of the collective momenta in the collective Hamiltonian is taken into account.
As a first attempt, a spin-dependent moment of inertia with the form of the $ab$-formula has been introduced into the collective
Hamiltonian, and such a calculation is referred as the 5DCH$^*$. The experimental energy spectra, the spin-rotational frequency relation,
the moments of inertia, as well as the ratios of $B(E2)/\mathcal{J}(I)$ of the yrast band in $^{102}$Pd are well reproduced by the
5DCH$^*$ calculations. That is, the 5DCH$^*$ could well describe the general characteristics of the tidal wave.

The shape evolution for the tidal wave in $^{102}$Pd is analysed in the framework of the 5DCH-CDFT by giving the average quadrupole
deformation parameters and the probability distributions of collective wave functions. It is found that the average deformation
$\langle\beta\rangle$ calculated by the 5DCH and 5DCH$^*$ both increase linearly with spin, while the increase obtained by 5DCH is
more obvious than that by the 5DCH$^*$. As analysed, this comes from the competition between the rotational kinetic energy
$\hat{T}_{\rm rot}$ and the collective potential $V$ and provides a schematic picture of the tidal wave.

In the present paper, by replacing the spin-independent MOIs with spin dependent MOIs, we effectively take the fourth order of the
collective momentum into account and well describe the tidal wave in $^{102}$Pd with the collective Hamiltonian.
One may argue that the MOIs calculated with the Thouless-Valatin  terms~\cite{libert1999microscopic} should be adopted rather than the
Inglis-Belyaev formula, as the dependence of MOIs on deformation may become considerably complicated especially for nuclei with soft potential
energy surfaces~\cite{prochniak2004self,prochniak2009quadrupole,hinohara2010microscopic,hinohara2012effect}. Such studies are
undoubtedly meaningful, which can answer whether the spin dependent MOIs in the collective Hamiltonian are still necessary or
not in this case. It is also noted that the mass parameters are just given from the cranking formula without any adjusted parameters.
Although further consideration of Thouless-Valatin corrections might somewhat change the excited energies~\cite{libert1999microscopic,yuldashbaeva1999mass},
its influence on the behavior of the yrast band would be rather small. Anyway, this paper has presented a possibility that the higher order effect
of collective momenta should be taken into account in the collective Hamiltonian.

\begin{acknowledgments}
The authors are indebted to Jie Meng for constructive guidance and valuable suggestions.
Fruitful discussions with Fangqi Chen, Umesh Garg, Tamara Nik{\v{s}}i{\'c}, Peter Ring, Dario Vretenar,
Shipeng Xie, Zhenhua Zhang, and Pengwei Zhao are highly acknowledged. This work is partly supported by
the Natural Science Foundation of China under Grants No. 11335002, No. 11375015, No.
11461141002, No. 11175002, No. 11235002, and No. 11005004, the Chinese Major State
973 Program No. 2013CB834400, and the China Postdoctoral Science Foundation under Grant
No. 2015M580007.
\end{acknowledgments}

\end{CJK*}
\end{document}